\documentclass{article}

\usepackage{arxiv}

\usepackage[utf8]{inputenc} 
\usepackage[T1]{fontenc}    
\usepackage{hyperref}       
\usepackage{amsmath}
\usepackage{amssymb}
\usepackage{graphicx}

\title{Maximum likelihood reconstruction of water Cherenkov events with deep generative neural networks}

\date{}

\author{Mo Jia, Karan Kumar, Michael J.~Wilking\\
  Department of Physics and Astronomy\\
  Stony Brook University\\
  Stony Brook, NY, U.S.A\\
  \And
  Liam S.~Mackey\\
  Department of Physics, Applied Physics and Astronomy \\
  Rensselaer Polytechnic Institute\\
  Troy, NY, U.S.A.\\
  \And
  Alexander Putra\\
  Department of Mathematics\\
  BMCC/City University of New York\\
  New York, NY, U.S.A.\\
  \And
  Cristovao Vilela\thanks{c.vilela@cern.ch}\\
  CERN European Organization for Nuclear Research\\
  CH-1211 Gen\`{e}ve 23, Switzerland\\
  \And
  Junjie Xia\\
  University of Tokyo\\
  Institute for Cosmic Ray Research\\
  Research Center for Cosmic Neutrinos\\
  5-1-5 Kashiwanoha, Kashiwa, Chiba, Japan\\
  \And
  Chiaki Yanagisawa\thanks{chiaki.yanagisawa@stonybrook.edu}\\
  Department of Physics and Astronomy\\
  Stony Brook University\\
  Stony Brook, NY, U.S.A\\
  Department of Science\\
  BMCC/City University of New York\\
  New York, NY, U.S.A.\\
  \And
  Karan Yang\\
  Information Science\\
  Cornell Tech\\
  New York, NY, U.S.A\\
}

\renewcommand{\headeright}{}
\renewcommand{\undertitle}{}
\renewcommand{\shorttitle}{}

\hypersetup{
pdftitle={Maximum likelihood reconstruction of water Cherenkov events with deep generative neural networks},
pdfsubject={hep-ex},
pdfauthor={Mo Jia, Karan Kumar, Liam S. Mackey, Alexander Putra, Cristovao Vilela, Michael J.~Wilking, Junjie Xia, Chiaki Yanagisawa, Karan Yang},
pdfkeywords={experimental particle physics, event reconstruction, water Cherenkov detectors, generative models, convolutional neural network},
}

\begin{document}
\maketitle

\begin{abstract}
Large water Cherenkov detectors have shaped our current knowledge of neutrino physics and nucleon decay, and will continue to do so in the foreseeable future. These highly capable detectors allow for directional and topological, as well as calorimetric information to be extracted from signals on their photosensors. The current state-of-the-art approach to water Cherenkov reconstruction relies on maximum-likelihood estimation, with several simplifying assumptions employed to make the problem tractable. In this paper, we describe neural networks that produce probability density functions for the signals at each photosensor, given a set of inputs that characterizes a particle in the detector. The neural networks we propose allow for likelihood-based approaches to event reconstruction with significantly fewer assumptions compared to traditional methods, and are thus expected to improve on the current performance of water Cherenkov detectors.
\end{abstract}


\section{Introduction}

In high energy physics several large water Cherenkov detectors have been used since 1980s such as IMB (Irvine-Michigan-Brookhaven)~\cite{IMB3}, Kamiokande (Kamioka Nucleon Decay)~\cite{Kamiokande}, Super-Kamiokande (SK)~\cite{Super-Kamiokande}, and for the near future Hyper-Kamiokande (HK)~\cite{HK} and proposed THEIA~\cite{THEIA} and ESSnuSB~\cite{ESSnuSB:2021azq} experiments. These are a type of detector that uses Cherenkov radiation produced by charged particles traveling faster than the speed of light in water. Photons in radiation traverse on a conical surface with its axis in the direction of parent charged particle. Photomultiplier tubes (PMTs) mounted on the walls of the detector detect these photons. PMTs produce electric charges proportional to the number of photons detected. The amount of charges and the arrival times are digitally recorded by electronic circuits. Depending on the direction of the parent charged particle with respect to the plane of PMTs, a detected Cherenkov ring leaves a pattern in the shape of a circle, an ellipse, or a parabola. For ultrarelativistic particles in water, Cherenkov photons are emitted at an angle of approximately 42$^\circ$ with respect to the particle direction. This angle decreases as particles slow down due to energy losses in the water. A good illustration of an atmospheric neutrino detected by the SK detector is available in the SK official site~\footnote{http://www-sk.icrr.u-tokyo.ac.jp/sk/detector/cherenkov-e.html}.  Water Cherenkov detectors contributed to the first detection of neutrinos from a supernova~\cite{Supernova}, the discovery of neutrino oscillation in atmospheric neutrinos~\cite{NuOsc}, the confirmation of solar neutrino oscillation~\cite{SolarNu} and search for proton decays~\cite{PDK}.

Neutrino and nucleon decay experiments pose a particular set of event reconstruction challenges. In order to overcome the smallness of neutrino cross sections and long nucleon lifetimes, these detectors are designed to have as much active mass as possible and, unlike typical collider or fixed-target experiments, the location of the events within the detector is not known, even approximately, \emph{a priori}. This challenge is compounded by a rich neutrino-nucleus interaction phenomenology~\cite{NuSTEC:2017hzk} at the energies of interest for many such experiments, in the order of GeV, which makes the detailed reconstruction of the events' final-state topology crucial both for precision measurements and potential discoveries.

Machine learning (ML) techniques have been increasingly adopted~\cite{Radovic:2018dip} to tackle these challenges across several detector technologies, from segmented scintillator detectors~\cite{Aurisano:2016jvx, MINERvA:2018smv, Alonso-Monsalve:2020nde} to liquid argon time-projection chambers~\cite{DUNE:2020gpm, MicroBooNE:2020hho}. Most of these efforts have focused on the \emph{discriminative} aspect of ML, with algorithms designed to classify events, for example into signal and background categories, or to estimate a variable of interest, such as the energy of an interacting neutrino.

In this work, we investigate a complementary approach by developing a ML-based \emph{generative} model that encodes the likelihood for the measurements at each PMT as a function of variables describing the event. This likelihood function can then be used to reconstruct events, for example using gradient-descent methods to find the event hypothesis that maximizes the likelihood, or by sampling the likelihood with Markov Chain Monte Carlo methods. This approach emphasizes interpretability by allowing for the detailed examination of the likelihood surface for each event, while guaranteeing powerful discrimination between competing event hypotheses by virtue of the Neyman-Pearson lemma~\cite{Neyman:1933wgr}.

\section{Water Cherenkov event reconstruction}
An event in a water Cherenkov detector consists of a set of charges and times recorded by each PMT. PMTs for which the amount of charge collected exceeds a given threshold will produce a hit, with a respective charge and time. An event is often broken into different hit clusters in time (subevents) and most read-out electronics systems effectively limit each PMT to measure a single charge (integrated over a period of time) and time (usually the time at which the electronic signal crosses the hit threshold) per subevent.  While these detectors can in principle have arbitrary shapes, most currently running and proposed experiments are cylindrical, with an instrumented barrel, top and bottom end-caps. Our work so far focuses on such cylindrical geometries, though it can in principle be adapted to other detector shapes.

The state-of-the-art in water Cherenkov event reconstruction is the FiTQun maximum likelihood estimation algorithm, whose adoption has led to improved neutrino oscillation measurements by long-running experiments~\cite{Super-Kamiokande:2019gzr,T2K:2019bcf}. At the core of FiTQun is a likelihood function which is evaluated over every PMT in the detector, regardless of whether or not it registers a hit in the event:
\begin{equation}
    L(x) = \prod_{j}^{unhit}P_{j}(unhit|x) \prod_{i}^{hit}\left\{1-P_{i}(unhit|x)\right\}f_{q}(q_{i}|x) f_{t}(t_{i}|x) \, ,
\label{eq:fqllh}
\end{equation}
\par
\noindent where $x$ denotes a set of parameters describing one or more particles in the detector, namely their type, starting positions, directions and momenta. The index $j$ runs over all the PMTs that did not register a hit and the index $i$ runs over all the hit PMTs, with $P_{j}(unhit|x)$ being the probability of a certain PMT not registering a hit under the $x$ hypothesis. The probability density function (hereafter PDF) for the observed charge $q_{i}$ in the $i^{\textrm{th}}$ hit PMT is $f_{q}(q_{i}|x)$, while the PDF for the observed time $t_{i}$ for the same PMT is $f_{t}(t_{i}|x)$.

Event reconstruction proceeds by minimizing the negative log-likelihood $-\ln{L(x)}$ by using the MINUIT gradient-descent package~\cite{James:1994vla}, with the hypothesis $x$ which minimizes $-\ln{L(x)}$ taken as the best-fit hypothesis for the event. This process is repeated for elements of $x$ which are categorical in nature, such as the particle type, or the number of particles (and their types) in the event. Likelihood ratio tests are used to discriminate between competing categorical hypotheses. In order to make this likelihood function tractable, it is factorized into several low-dimension components. In particular, the PDFs associated to Cherenkov photons that produce a hit without having scattered in the water or reflected in the detector surfaces are factorized from the PDFs that describe so-called indirect photons that scatter or reflect before producing a PMT hit. The level of detail of the indirect photon PDF is limited by the high number of dimensions required for it to be fully specified. In particular, this limitation makes it difficult to reconstruct heavier, typically slower, particles such as protons, since it is challenging to accommodate the effect of the decreasing Cherenkov photon emission angle. Finally, the each component of the factorized likelihood needs to be tuned separately to the detector geometry of interest, requiring a large amount of bespoke simulated data with different components of the simulation disabled in turn.

\section{Generative neural networks for maximum likelihood reconstruction}
In order to overcome the challenges in water Cherenkov event reconstruction associated to the curse of dimensionality, we have designed a convolutional generative neural network to replace the factorized likelihood function in FiTQun. The output of this network are PDFs for the hit charges and times at each PMT in the detector. Given the success of ML methods in processing high dimensional data, we do not factorize the likelihood function in our approach, nor do we require bespoke sets of simulated data for training. Rather, the neural network can be trained using a regular, fully detailed, simulated data set.

Like in the existing FiTQun algorithm, the likelihood function for multiple particles in an event can in principle be combined to form complex event hypotheses. While this capability is a future goal of our project, in the present work we have focused on demonstrating the method using single-particle events consisting of either a showering electron, or a track-like muon.

\subsection{Network architectures}
\begin{figure}[!ht]
  \begin{center}
    \includegraphics[width=\textwidth]{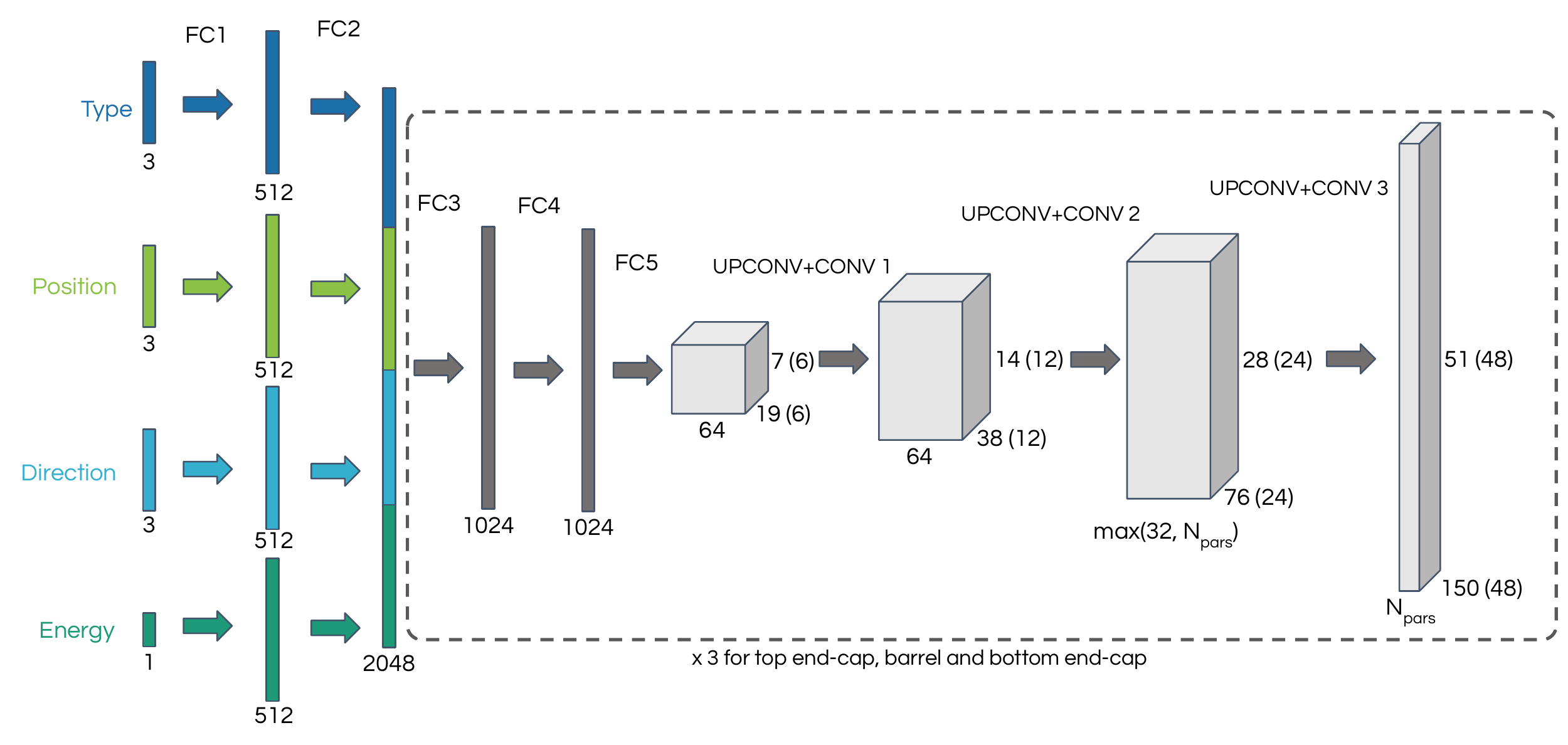}
  \end{center}
  \caption{Generative neural network architecture diagram. The input describes a single-particle state and consists of a one-hot vector encoding the particle type, two three-dimensional vectors corresponding to the particle starting position and direction cosines, and the particle energy. The section of the network enclosed in a dashed line is repeated three times, one for each region of the detector: two end-caps and one cylindrical barrel. The dimensions in brackets correspond to the end-caps. The number of output channels, $\textsf{N}_{\textsf{pars}}$, depends on the parameterization of the loss function.}
  \label{fig:architecture}
\end{figure}

We have implemented our model using the \emph{PyTorch}~\cite{pytorch_ref} framework, and we used the network architecture in \cite{DBLP:tablechaircar} as a starting point. As depicted in Figure~\ref{fig:architecture}, our network is composed of two parts: five fully connected (FC) layers are followed by three pairs of transposed convolution (UPCONV) and convolution (CONV) layers. We have found that pairing the two types of convolutional layers as done in \cite{DBLP:tablechaircar} results in smoother outputs. However, as discussed below this approach may also limit the network's ability to reproduce sharp features in our data. \textit{ReLU} activation is used after each layer.

The neural network input is a vector describing a single-particle state, consisting of a three-dimensional one-hot encoding of the particle type (electron, muon or gamma\footnote{We do not train with gammas in this iteration, but instead include a place holder for it.}), a three-dimensional vector with the particle starting position, a three-dimensional vector of the particle direction cosines and finally a scalar corresponding to the particle energy. Following the architecture in \cite{DBLP:tablechaircar}, each input type is processed through two fully connected layers of 512 nodes each on its own (FC1 and FC2). The output is concatenated into a feature vector of length 2048 which passes through two fully connected layers of 1024 nodes (FC3 and FC4) followed by a final fully connected layer (FC5) which results in the starting point for the convolutional part of the network. The three convolutional layers result in images where each pixel represents a PMT in the detector and the pixel values encode parameters which are used to build the likelihood function.

The portion of the network starting with FC3 is repeated three times, one for each section of the cylindrical detector: the top and bottom end-caps, and the cylindrical barrel. Each of these detector regions is represented as two-dimensional images with $48 \times 48$ pixels for the end-caps and $150 \times 51$ pixels for the barrel. The depth of the output depends on the parameterization of the likelihood function and ranges from four in the simplest case to 61 in the most complex case.

A kernel size of $4 \times 4$ and a stride of $2 \times 2$ is used for all UPCONV layers and the CONV layers use a $3 \times 3$ kernel and $1 \times 1$ stride. The first UPCONV + CONV pair layer takes a tensor of depth 64 and produces an output of depth of 64, and the second layer reduces the depth to either 32 or the number of output channels, whichever is largest. The final layer produces the desired number of output channels, which depends on the choice of loss function. Padding is used on the output to match the odd dimension of the detector barrel.

To match the square images produced by the neural network to the circular end-caps of the detector, the pixels close to the corners, which do not correspond to physical PMTs, are masked when evaluating the loss function.

\subsection{Loss functions}
\label{sec:lossfunction}
We have designed the loss function in a similar fashion to Eq.(\ref{eq:fqllh}), with two components: one describing the probability of the PMT being hit and the other describing the probability density function for a hit charge and time:
\begin{equation}
  \textrm{Loss} = -\log L = \sum_i - \log P_{unhit}(y_i) + \sum_{i_{hit}} -\log p_{qt}(q_{i_{hit}}, t_{i_{hit}}) \, ,
  \label{eq:cringellh}
\end{equation}
\noindent where the index $i$ runs over all the PMTs in the detector, $y_i$ is a label set to 1 if the PMT is not hit or 0 if the PMT is hit, $i_{hit}$ runs only over the PMTs which are hit in the event and $p_{qt}(q_{i_{hit}}, t_{i_{hit}})$ is the PDF for observing charge $q_{i_{hit}}$ and time $t_{i_{hit}}$. The loss function is the sum of the negative log-likelihood over all PMTs in the three regions of the detector.

For the PMT hit probability we use the \emph{PyTorch} built-in function \texttt{BCEWithLogitsLoss} which implements the binary cross-entropy loss~\cite{cover1991infotheory_elements} (equivalent to the negative log-likelihood) using a \textit{Sigmoid} function to regularize predictions of hit probability. A single channel of the neural network output represents the logit of the hit probability.

\begin{figure}[ht!]
\begin{center}
\includegraphics[width=\textwidth]{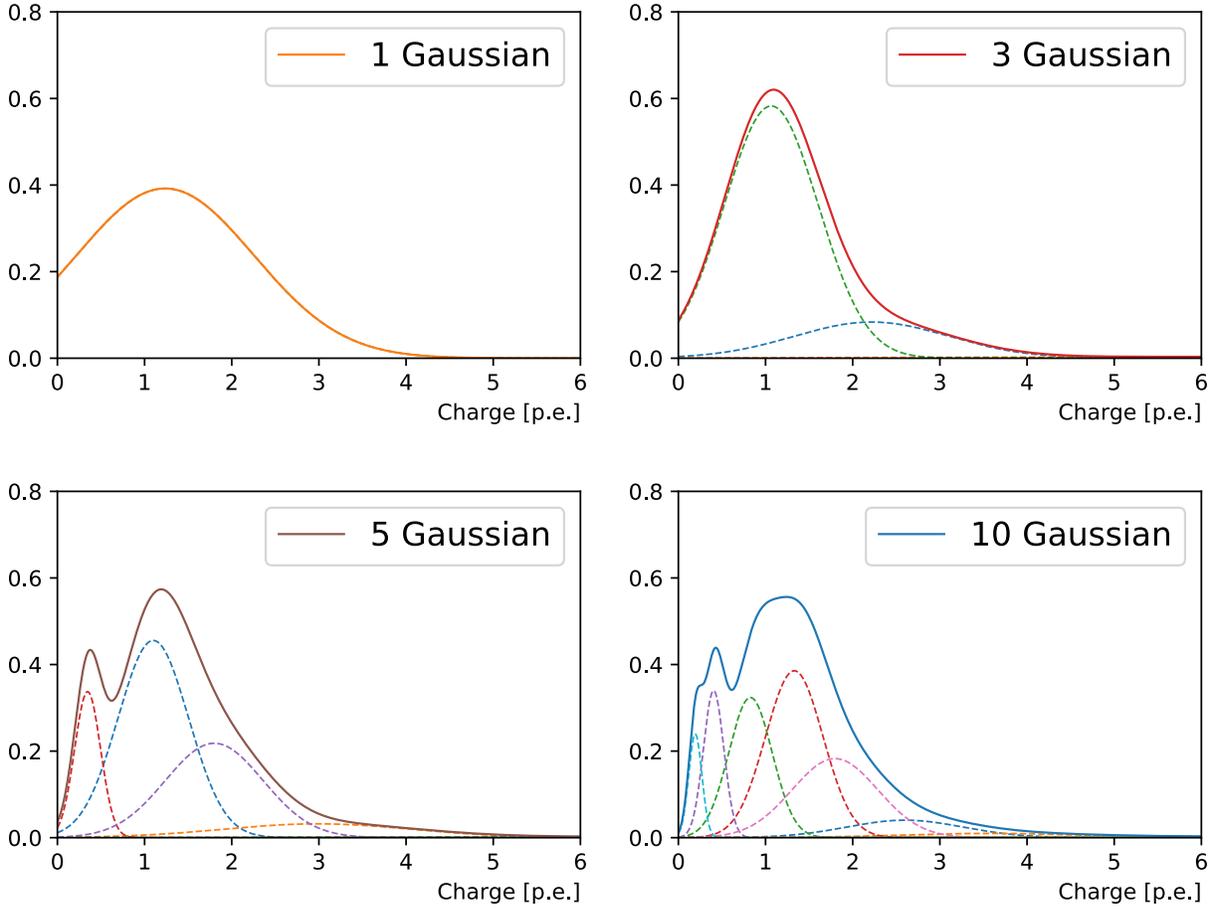}
\end{center}
\caption{Charge response PDF of a randomly chosen PMT in a 1 GeV muon event originating from the detector center and propagating in $+x$ direction. From the top left to bottom right panel are the reconstructions of 1-Gaussian, 3-Gaussian, 5-Gaussian, and 10-Gaussian charge-only network respectively. With the increasing number of sub-components, we first observe the separation of the narrow peak from the prompt hits and the broad tail possibly associated to the delayed ones. Then a sharp peak at small charge is added. With even more sub-components both peaks will start broadening.}\label{fig:multigaus_components}
\end{figure}

It was observed in \cite{pd18:hkpmtcalib} and this work that both the charge and timing of a PMT's responses can be very non-Gaussian, despite the stochastic process of photo-electron multiplication, due to a list of reasons ranging from the kinematics of event to the reflection of light from ambient environment. In order to accommodate the \emph{a priori} unknown functional form of the hit charge and time PDF, we approximate this function with a weighted mixture of Gaussian PDFs in one dimension (taking into account only the hit charge) or two dimensions (hit charge and time). Figure~\ref{fig:multigaus_components} shows an example of the charge PDF predicted by networks with different numbers of Gaussian components for an arbitrarily chosen PMT.

We have explored several combinations of the number of components in the Gaussian mixture in one and two dimensions, and different parameterizations for the two-dimensional Gaussian function, either keeping the charge and time uncorrelated, or including a correlation factor. Each component in the Gaussian mixture is weighted by a coefficient which corresponds to one of the neural network output channels. To preserve the normalization of the PDFs a \textit{Softmax} function is applied to the coefficients. 

In the case where the hit charge and time are treated as uncorrelated in each of the Gaussian components, the mixture of Gaussians PDF is given by:
\begin{multline}
  \label{eq:qtlosssum}
  -\log p_{qt}(q_{i_{hit}}, t_{i_{hit}}) = -\sum_{i_{hit}}\Bigg[\sum^{N}_{j} \Bigg(\log(n_{j}) - \log({\sqrt{2\pi}\sigma_{q_{j}}}) - \frac{(q_{i_{hit}}-\mu_{q_{j}})^{2}}{2\sigma_{q_{j}}^2} \\ - \log({\sqrt{2\pi}\sigma_{t_{j}}}) -\frac{(t_{i_{hit}}-\mu_{t_{j}})^{2}}{2\sigma_{t_{j}}^2}\Bigg)\Bigg]_{i_{hit}} \, ,
\end{multline}
\noindent where $i_{hit}$ runs over the hit PMTs, as in Eq.(\ref{eq:cringellh}), $j$ runs over the $N$ Gaussian components, $n_j$ is the normalization factor for the $j$-th component, $\mu_{q_{j}}$ and $\mu_{t_{j}}$ are the charge and time means for the $j$-th component, respectively, and $\sigma_{q_{j}}$ and $\sigma_{t_{j}}$ are the corresponding standard deviations. For the one-dimensional case, where only the hit charges are considered but not the times, the second line of the equation is omitted. In order to improve numerical stability, \emph{PyTorch}'s implementation of the log-sum-exp function is used to evaluate the loss function.

For each component in the Gaussian mixture, a set of network output channels correspond to $\log \mu_{q_{j}}$, $\log \sigma_{q_{j}}$, $\mu_{t_{j}}$, and $\log \sigma_{t_{j}}$, with the logarithms being used to ensure non-negative values of the hit charge and standard deviations. Together with the hit probability and the Gaussian component coefficients, the total number of channels is $1 + N \times 3$ for the one-dimensional case and $1 + N \times 5$ for the two-dimensional case.

In the bivariate case, correlations between the charge and time dimensions of the loss function above can arise from the independent sets of $\mu_{q_{j}}$ and $\mu_{t_{j}}$. However, the most general form of the two-dimensional mixture of Gaussians includes a covariance term describing the correlation between charge and time within each of the mixture's components. To realize this goal we formulate two-dimensional PDF of correlated charge and time responses in the way introduced by \cite*{P.M.Williams}:
\begin{equation}
\label{eq:qtcorrLH}
    f(\,\boldsymbol{\eta}|\,\boldsymbol{\theta}) = \sum_{j}^{N} \frac{n_j}{(2\pi)|\Sigma_{j}|^{1/2}}\exp{-\frac{1}{2}(\boldsymbol{\eta} - \boldsymbol{\theta_j})^{\mathrm{T}}\Sigma_{j}^{-1}(\boldsymbol{\eta} - \boldsymbol{\theta_j})} \, ,
\end{equation}

\noindent where $\boldsymbol{\eta}$ and $\boldsymbol{\theta}$ include both the charge and time as a two-dimensional vector: 
\begin{align}
    \boldsymbol{\eta} &= (t, q) \\
    \boldsymbol{\theta_j} &= (\mu_t, \mu_q)_j
\end{align}
\noindent where $\Sigma_i$ is a two-dimensional covariance matrix.

To improve numerical stability we use triangular matrix instead of the full covariance matrix $\Sigma$ by Cholesky decomposition:
\begin{equation}
    \Sigma^{-1} = \begin{pmatrix}
    \alpha_{11} & 0 \\
    \alpha_{12} & \alpha_{22}
    \end{pmatrix}
    \begin{pmatrix}
    \alpha_{11} & \alpha_{12} \\
    0 & \alpha_{22}
    \end{pmatrix}
\end{equation}
with positive-only diagonal terms and
\begin{equation}
    |\Sigma|^{-1/2} = \alpha_{11}\alpha_{22}
\end{equation}

The $\alpha$'s are also predicted by the neural network, requiring a total of $1 + 6\times N$ output channels. To ensure positive definite matrix $\Sigma$, we keep the absolute value of $\alpha_{12}$ while forcing its sign so that $(\eta - \theta_i)^{\textrm{T}}\Sigma_{i}^{-1}(\eta - \theta_i)>0$ for $\forall{\eta,\, \theta}$.

\[ 
\alpha_{12} \left\{ \begin{array}{l} < 0,\,\,\, \mathrm{if}\,\,\, (t - \mu_t)(q - \mu_q) < 0 \\ > 0,\,\,\, \mathrm{if}\,\,\, (t - \mu_t)(q - \mu_q) > 0
\end{array} \right.
\]

\section{Data sets}
\label{sec:sksample}
The data set used in this work consists of single-electron and single-muon events generated with random kinematics and positions in a model of the SK detector --- a cylindrical volume with 36 meters in height and 34 meters in diameter. The coordinate system used to describe the detector has $z$ pointing along the cylinder axis. The data are simulated with WCSim, an open-source program based on Geant4~\cite{GEANT4:2002zbu} and ROOT~\cite{Brun:1997pa}, which models the particle propagation in the detector and the electronic response of the PMTs. For each simulated event, its Monte Carlo truth, including particle type, position, direction cosines and energy, is saved in order to be used during neural network training. The hit charges and times at every PMT in each event are stored in three two-dimensional arrays representing the unrolled cylindrical barrel of the detector ($151 \times 50$) and the two circular end-caps ($48 \times 48$). The data set consists of 1,003,200 electron events and 1,049,415 muon events. The kinetic energy of the electrons is uniformly distributed between 1 MeV to 6500 MeV and muons between 150 MeV to 6500 MeV. Spontaneous discharges in the PMTs, a source of uncorrelated noise commonly referred to as the dark rate, is not simulated, for simplicity. Taking into account the PMTs' dark rate is a future goal of this project. In order to produce pure single-particle data sets, delayed electrons originating from muon decays are not simulated.

In addition to the training data set, we have also produced another set of muon and electron events with fixed position, direction, and energy. These events originate in the center of the detector and the particles propagate along the $x$ direction (onto the cylinder barrel) with a kinetic energy of 500 MeV. These events are used to evaluate the quality of the likelihood functions generated by the neural networks by comparing them to the hit probability extracted directly from the simulation as well as the distributions of simulated hit charges and times.

\section{Training}
\begin{figure}[h!]
\begin{center}
\includegraphics[width=\textwidth]{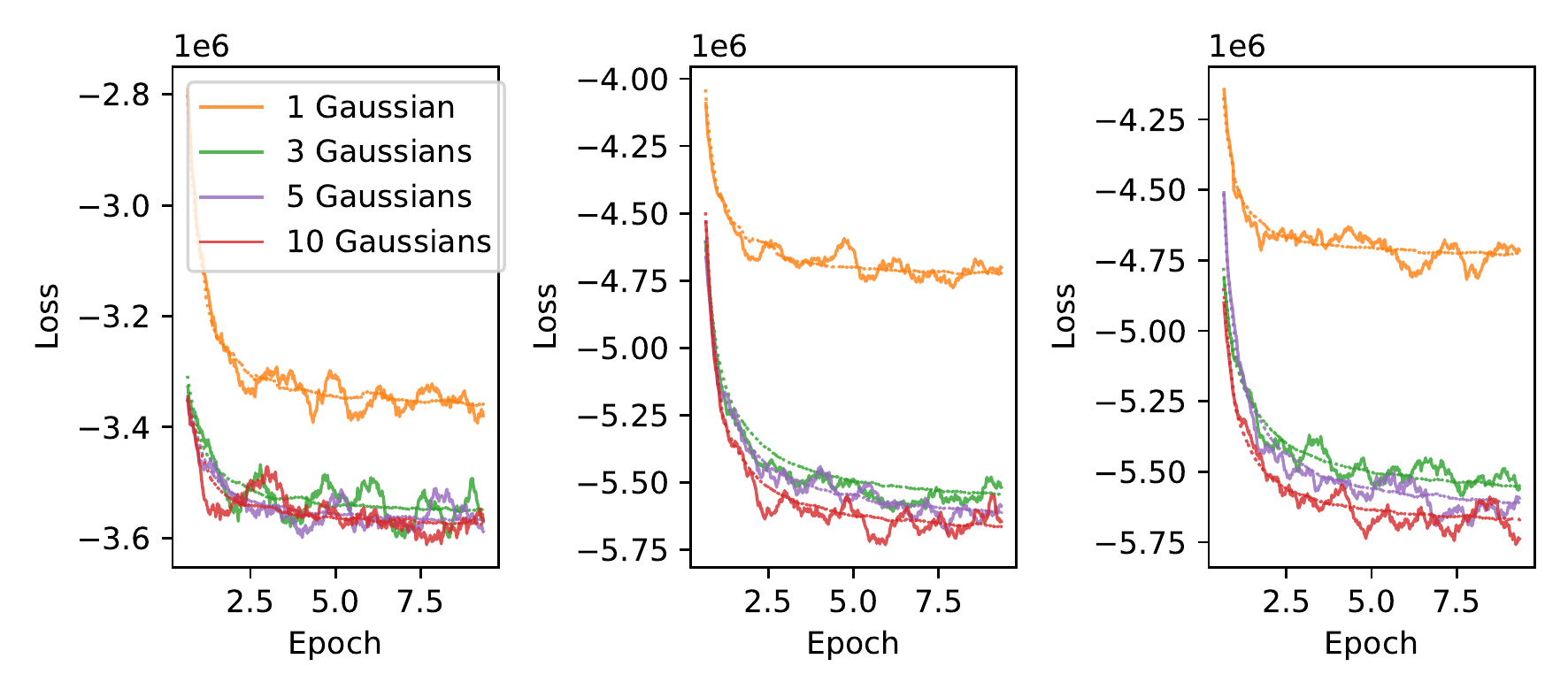}
\end{center}
\caption{Neural network training curves. The loss is shown as a function of training epoch for 1, 3, 5 and 10 component charge-only (left), uncorrelated charge and time (middle), and correlated charge and time likelihoods. The solid lines correspond to the loss evaluated on a validation sample, and the dotted lines correspond to the loss evaluated using the training sample.}
\label{fig:trainingcurves}
\end{figure}

We use 75\% of the randomized electron and muon events for training, with the remaining 25\% used for \emph{in-situ} validation every 100 iterations. The network is trained using minibatches of 200 shuffled events. The results presented in this work were obtained by training the neural networks for 10 epochs, using the \texttt{Adam}~\cite{kingma2017adam} optimizer with the initial learning rate set to 0.0002, and all other parameters left at the \emph{PyTorch} default values. To improve numerical stability we normalize the particle position by the detector dimensions, the energy by 5000 MeV, PMT hit charges by 2500 p.e., and convert PMT hit times to $\mu$sec with an offset of -1 $\mu$sec so that the values are of $\mathcal{O}(10^{-3}\sim1)$. We found that this pre-normalization gave the training more stability compared to using \texttt{BatchNorm} layers on the inputs, and using both \texttt{BatchNorm} layers and input pre-normalization resulted in slower training. Figure~\ref{fig:trainingcurves} shows examples of training curves for different configurations of the neural network. The configurations with a single Gaussian component or one-dimensional loss function generally converge within 10 epochs, while networks with multiple components or two-dimensional loss functions still show downward trends at the end of 10 epochs of training. Therefore, the results presented in this publication can likely be improved with extended training of the networks. We have saved both the trained network weights and optimizer states at 10 epochs, which can be used for further training beyond this point.

\section{Results}
For the neural networks described in this work to be effective in maximum likelihood reconstruction, they need to describe the data as closely as possible, to avoid bias, and they should be smooth to avoid local minima in the likelihood surface which would make event reconstruction challenging. In this section we examine the output of the neural network in two ways: first we compare the neural network prediction to the distributions obtained directly from the simulated data set produced with fixed particle kinematics; we then use the randomized validation data set to scan the likelihood function as a function of the particle energy to evaluate the smoothness, bias, and the accuracy of the neural network in identifying the true particle type in the events.

\subsection{Comparison of neural network output to events simulated with fixed kinematics}
\begin{figure}[h!]
\begin{center}
\includegraphics[width=\textwidth]{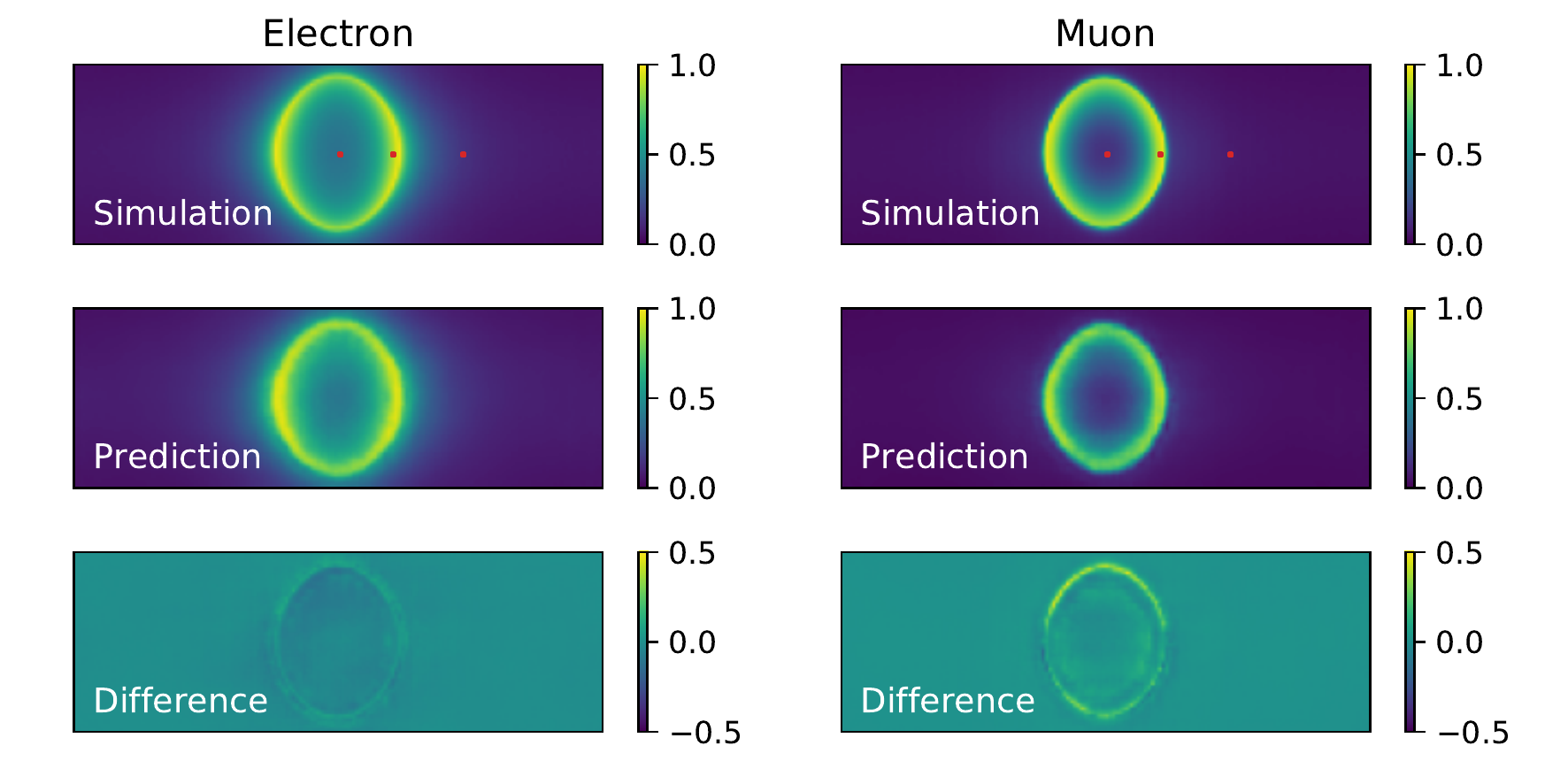}
\end{center}
\caption{Comparison of the hit probability for simulated (top) electron (left) and muon (right) events generated with fixed kinematics, and the neural network prediction (middle). The difference between the neural network prediction and the simulation is shown in the bottom. These results were produced with the single-component charge-only loss function. The red marks on the top panel figure indicate the location of the three reference PMTs chosen to examine the hit charge and time PDFs.}
\label{fig:fixed_hitprob}
\end{figure}

To evaluate the accuracy of the neural network at predicting the probability of PMTs in the detector being hit, we measure the hit probability in the simulation by counting the number of times each PMT is hit in the set of events with fixed kinematics and dividing each PMT's hit count by the total number of events in the set. The resulting hit probabilities are shown Figure~\ref{fig:fixed_hitprob}, where the Cherenkov ring pattern is clearly seen and the difference between the fuzzy rings produced by showering electrons and sharp rings produced by track-like muons is evident. For the electron events, the agreement between the neural network prediction (middle panel of the same figure) and the simulation is excellent. On the other hand, for muon events there is some level of residual difference (bottom panel) that might indicate the neural network might have limited ability to reproduce sharp features in the data, a shortcoming that is common in generative convolutional neural networks~\cite{durall2020upconv}. In both cases it is clear the neural network reproduces the general features of the events, including a clear difference between the fuzzy rings predicted for electrons and sharp rings for muons.

\label{sec:particlegun}

\begin{figure}[h!]
\begin{center}
\includegraphics[width=\textwidth]{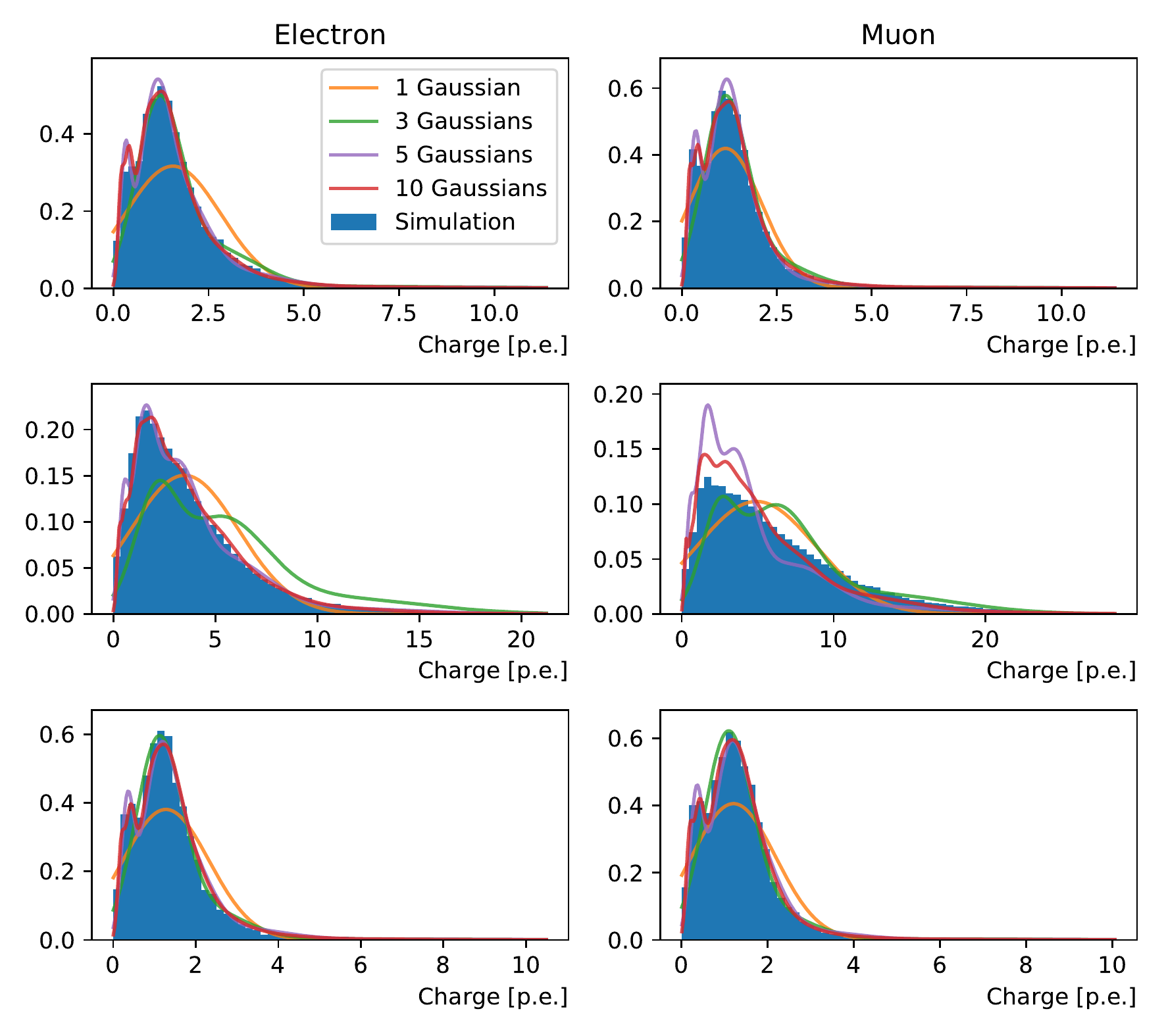}
\end{center}
\caption{Distributions of hit charges for electrons (left) and muons (right) events in the three reference PMTs: in the center (top), edge (middle) and, outside (bottom) of the Cherenkov ring pattern produced by events generated with fixed kinematics. Predictions of the neural network trained with one, three, five and ten Gaussian components are superimposed on the simulation.}
\label{fig:fixed_chargepdf}
\end{figure}

To inspect the PDFs for the hit charges and times we have chosen three reference PMTs in the detector, marked in red in Figure~\ref{fig:fixed_hitprob}, which are located in the center, edge, and outside of the Cherenkov rings produced by the events generated with fixed kinematics. The PDFs resulting from training the neural network with the charge-only, one-dimensional, loss function using one, three, five and ten components is shown in Figure~\ref{fig:fixed_chargepdf}. It is clearly seen that the distributions of simulated hit charges are not Gaussian and therefore the single-Gaussian PDF describes the data very poorly. With three components, the neural network is able to reproduce general features in the data, such as the high-charge tail present in all distributions. With five and particularly with ten components the PDFs describe the distributions in detail, including a slight bi-modality seen at low charge. We have observed that the neural network struggles to reproduce the charge distribution for PMTs on the sharp edge of the muon ring. Given the shape of this distribution is unremarkable compared to all others, we believe this artifact might be due to the known shortcomings of convolutional generative networks to reproduce sharp features in the data, as identified in the hit probability prediction above.

\begin{figure}[h!]
\begin{center}
\includegraphics[width=\textwidth]{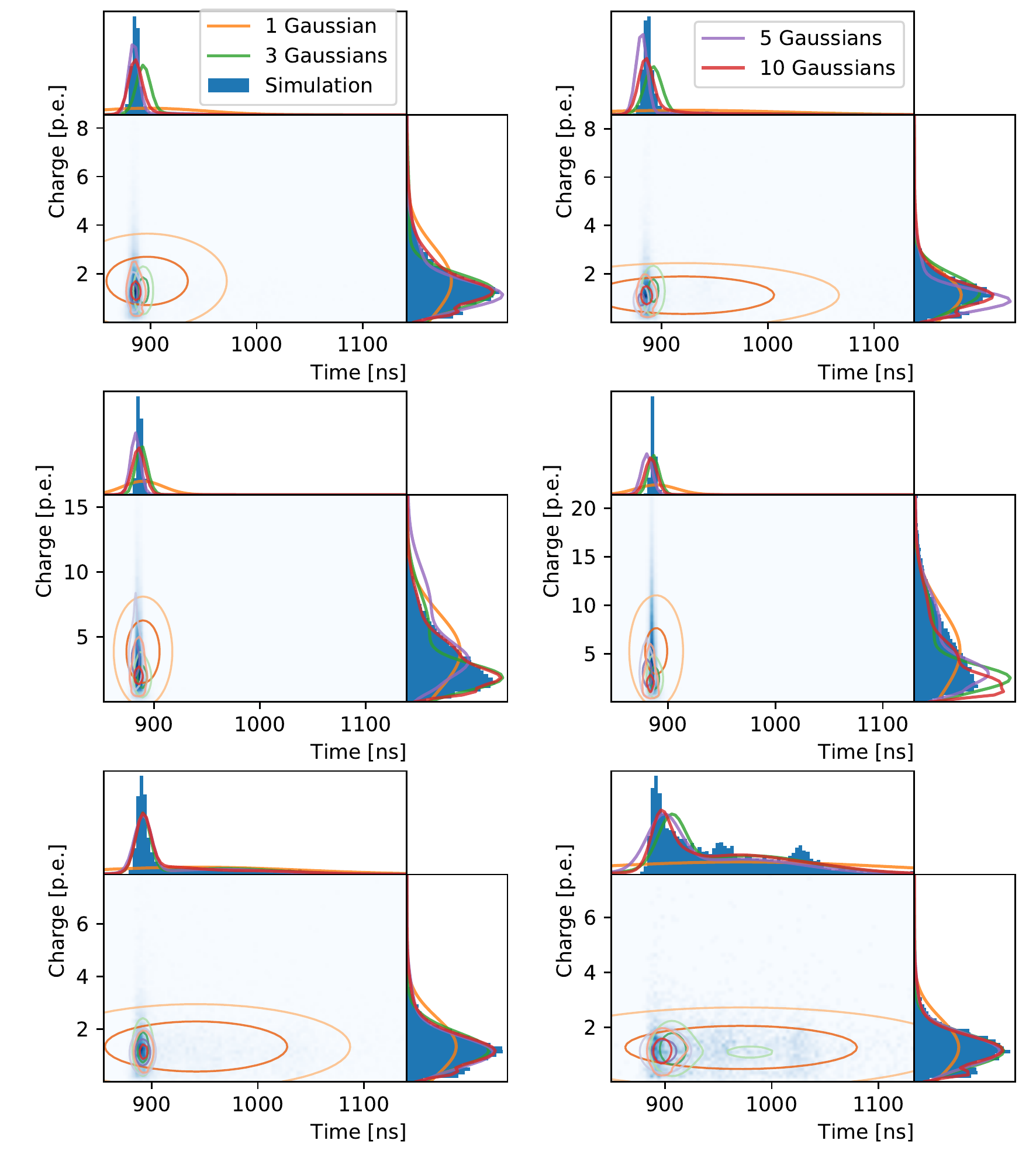}
\end{center}
\caption{Distributions of hit charges and times for electrons (left) and muons (right) events in the three reference PMTs: in the center (top), edge (middle) and, outside (bottom) of the Cherenkov ring. The two-dimensional distributions are shown with time on the abscissa and charge on the ordinate together with their one-dimensional projections and the neural network prediction.}
\label{fig:fixed_chargetimecorrpdf}
\end{figure}

The two-dimensional loss function using correlated Gaussian components is inspected similarly for the three chosen PMTs in Figure~\ref{fig:fixed_chargetimecorrpdf}. The inadequacy of a single two-dimensional Gaussian to describe the data is even more staggering than in the one-dimensional case. This is expected, given the time distribution is multi-modal, with a sharp peak associated to Cherenkov photons that produce hits without scattering in the water or reflecting off the detector surfaces, and a long tail of scattered and reflected photons, including hints of reflection peaks. The width of the sharp peak is mostly determined by the PMT response, while the distribution of late hits depends strongly on the detector geometry and water properties. A three-component model shows a much better fit, with the network reproducing the multi-modality in the time distribution. The five and ten component models produce increasingly complex shapes with correlations clearly seen in the time and charge dimensions. As in the previous two examples, the neural network's worst performance is for the PMT on the edge of the muon Cherenkov ring, with a bias in the time prediction observed in addition to the relatively poor agreement in the charge prediction.

\subsection{Likelihood scans as a function of particle energy}
\label{sec:likelihoodscan}
\begin{figure}[h!]
\begin{center}
\includegraphics[width=\textwidth]{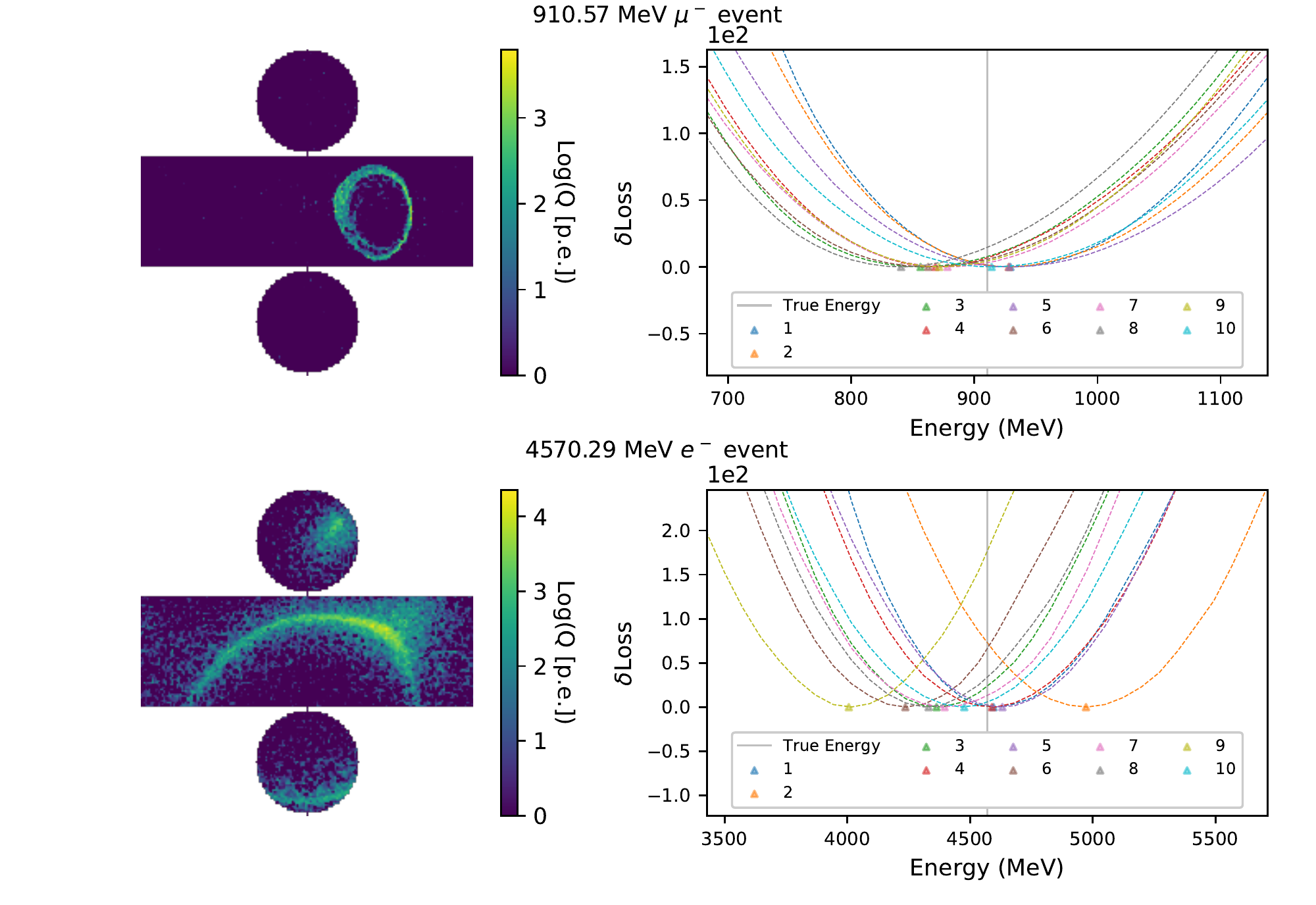}
\end{center}
\caption{Energy scans of a muon (electron) event are shown in the top (bottom) row. The left column shows the simulated hit charges in each event in logarithmic scale, with the cylindrical detector's surface unrolled in two dimensions. In the right column the y-axis shows $\delta{\mathrm{Loss}(q)}$ with the minimum in each curve subtracted. The energies that minimize the likelihood functions, $E_{rec}$, of charge-only networks with 1$\sim$10 Gaussian components are marked with triangles. The event's true energy is shown as a solid vertical line.}
\label{fig:eventdisp_8_multigaus_curves}
\end{figure}

For the method proposed in this work to be viable as an event reconstruction technique, it is important that the likelihood function encoded in the neural network is smooth, so that it can be used in gradient-descent algorithms, and that the minimum of the likelihood surface lies close to the true parameters describing the event. To examine these characteristics of the neural network, we scan the likelihood surface of the randomized data set used for training the neural network, with 100 uniform steps in energy ranging -80\% to +80\% of the event's true energy. The other neural network input parameters remain fixed to their true values. At each scan point, we replace the event's true energy with the shifted one and evaluate the neural network using this hypothesis as the input to calculate the corresponding loss. A quadruple spline interpolation is applied to the 100 scanned points to find the energy that minimizes the loss, or in other words maximizes the likelihood function. Figure~\ref{fig:eventdisp_8_multigaus_curves} presents an example of the interpolated energy scan from multi-Gaussian charge-only networks of intermediate energy muon and electron events. 

We take the energy which minimizes the loss function to be the estimator for the true particle energy, or the \emph{reconstructed} energy -- $E_{rec}$, and the fractional energy residual $\Delta_{E}$ defined in Eq.(\ref{eq:eres}) is used to measure the neural network's energy reconstruction performance.
\begin{equation}
\label{eq:eres}
    \Delta_{E} = \frac{E_{rec} - E_{true}}{E_{true}}
\end{equation}

\begin{figure}[h!]
\begin{center}
\includegraphics[width=\textwidth]{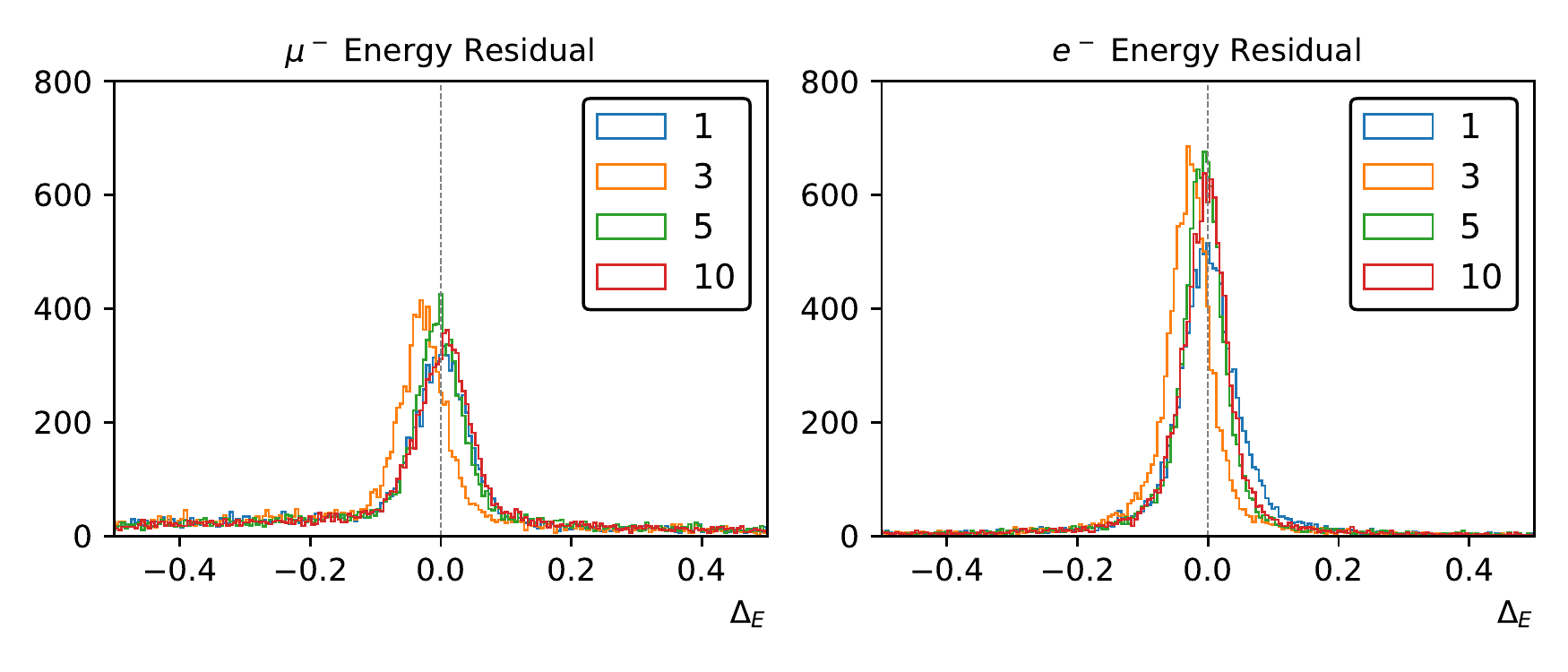}
\end{center}
\caption{Energy reconstruction performance for muons (left) and electrons (right) using the charge-only loss functions with one, three, five and ten Gaussian components.}
\label{fig:eresiduals}
\end{figure}

Figure~\ref{fig:eresiduals} shows the distribution of $\Delta_{E}$ for sets of 12000 muon and electron events, using various multi-Gaussian charge-only networks. Unbiased energy reconstruction is achieved with one, five and ten components, while the three-Gaussian loss function results in slightly biased energy reconstruction. Especially for electron events, a higher number of components results in improved energy reconstruction.

We have applied the likelihood scan to all the three loss functional forms described in Section \ref{sec:lossfunction} but we show results for the charge-only and charge-time including correlations in this section, because the two networks using PMT time responses have similar reconstruction performance and the correlated version is slightly better.

Under certain circumstances the network will face trouble reconstructing events, especially when not all of the particle energy is deposited inside the detector. These events are more likely to cause reconstruction failure as no valid minimum is found in the loss function scan. In general the reconstruction failure rate is $\leq$4\% for electron and $\leq$10\% for muon events. Both charge-only and charge-time correlated loss functions have better reconstruction performance for electron events, whereas in the charge-time correlated networks the fraction of unsuccessful scans increases with the number of loss function components.

\begin{figure}[h!]
\begin{center}
\includegraphics[width=\textwidth]{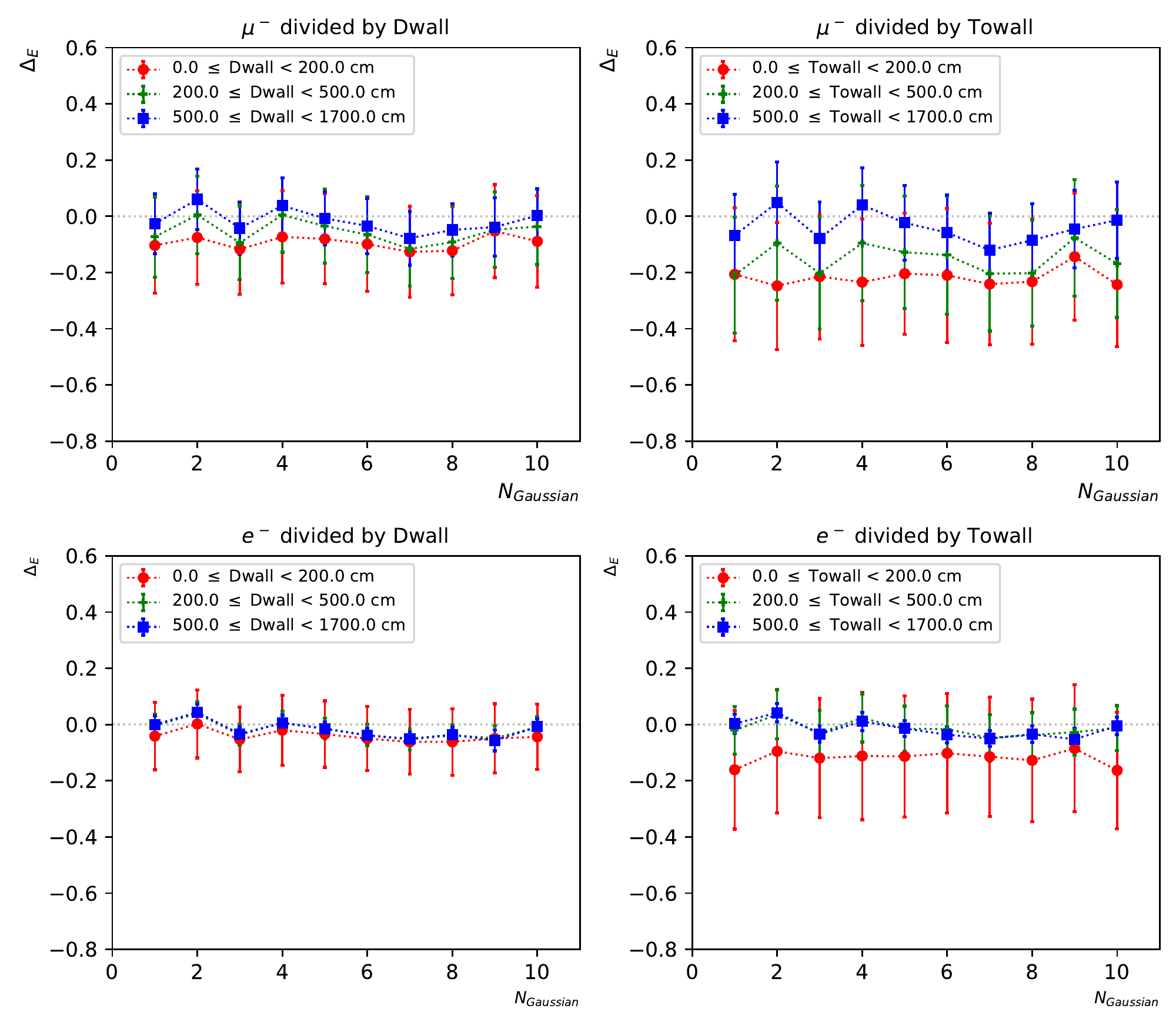}
\end{center}
\caption{The statistical mean $\Delta_{E}$ of charge-only networks with 1$\sim$10 Gaussian from muon and electron events respectively. The three regions are 0$\sim$200, 200$\sim$500, and $\geq
$500 cm from the detector walls. The left column shows dependence of ``dwall" and the right ``towall". Error bars represent the standard deviation of $\Delta_{E}$ in each event set, which can be viewed as the energy resolution. }
\label{fig:multigaus_ereso_vs_ngaus}
\end{figure}

\begin{figure}[h!]
\begin{center}
\includegraphics[width=\textwidth]{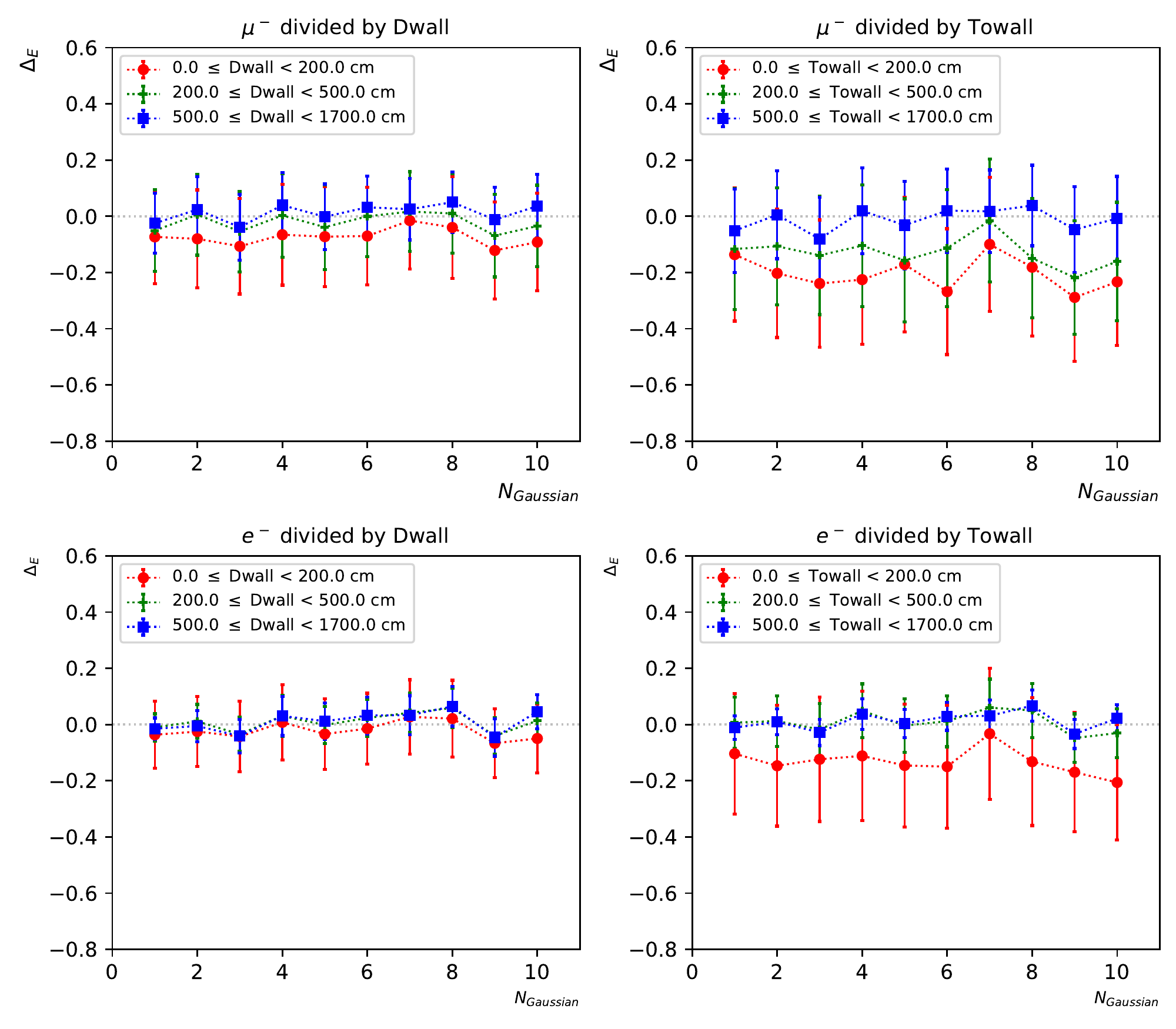}
\end{center}
\caption{The statistical mean $\Delta_{E}$ of charge-time correlated networks with 1$\sim$10 Gaussian from muon and electron events respectively. The three regions are 0$\sim$200, 200$\sim$500, and $\geq
$500 cm from the detector walls. The left column shows dependence of ``dwall" and the right ``towall". Error bars represent the standard deviation of $\Delta_{E}$ in each event set, which can be viewed as the energy resolution.}
\label{fig:multigaus_qtcorr_ereso_vs_ngaus}
\end{figure}

To further investigate the neural network reconstruction performance we introduce two parameters: ``dwall" is the distance between the particle starting position and the nearest detector wall, and ``towall" is the distance to the nearest detector wall along the particle's moving direction. For a particle with small ``dwall" and ``towall", it has higher probability to escape the detector and thus not all of the particle energy will be detected. Figure~\ref{fig:multigaus_ereso_vs_ngaus} and \ref{fig:multigaus_qtcorr_ereso_vs_ngaus} present the energy reconstruction performance of multi-Gaussian networks in the detector regions defined by the ``dwall" and ``towall" parameters. Events for which no minimum is found in the loss function scan are excluded. As expected the performance is worse for events with smaller ``towall", with the particle energy being systematically underestimated. The dependence of performance on ``towall" is more significant for muons, which tend to travel much longer distances before they drop below the Cherenkov threshold. Due to the longer tails in their $\Delta_{E}$ distributions, the muon energy resolution is worse across the three detector regions whereas the electron energy resolution is superior in the central region. For events sufficiently far from the the walls, the mean $E_{rec}$ predicted by any of the 1$\sim$10-Gaussian networks is unbiased, with fluctuations smaller than the standard deviation.


\subsubsection{Particle identification}
\label{sec:PID}
\begin{figure}[h!]
\begin{center}
\includegraphics[width=\textwidth]{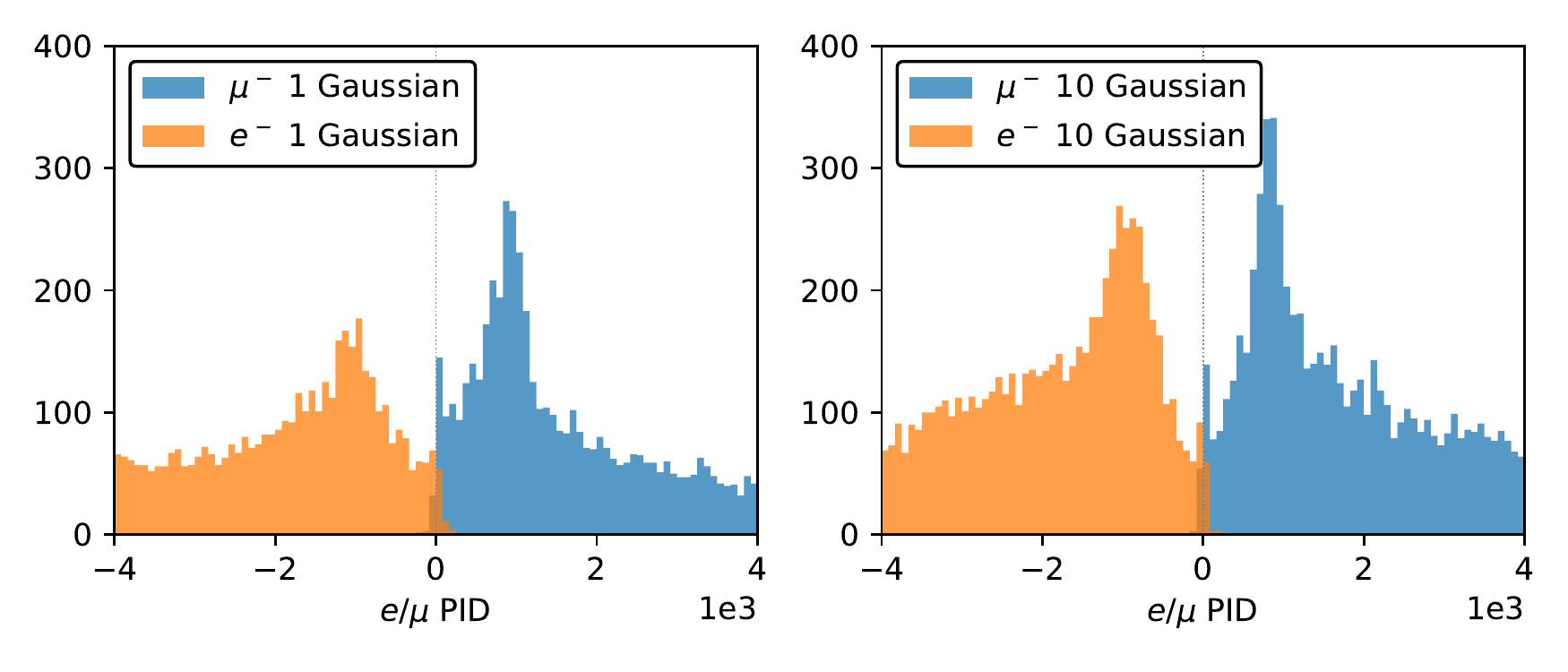}
\end{center}
\caption{PID performance of the one-Gaussian (left) and ten-Gaussian (right) charge-only networks. The true electron and muon events locate in the negative and positive region respectively, and the ten-Gaussian network shows better concentrated peaks for both particle types.}
\label{fig:SKPIDScan}
\end{figure}

The trained neural networks are able to separate electron and muon events by comparing the loss values of the competing hypotheses. To study the particle identification (PID) performance of the neural network we use the loss function scans described in Section~\ref{sec:likelihoodscan} above. For each event, we take the difference in the loss value at the energy that minimizes the loss, $E_{rec}$, while keeping all other input parameters fixed. The resulting variable takes negative values for electron-like events and positive values for muon-like events:
\begin{equation}
    e/\mu\,\,\mathrm{PID} = \mathrm{Loss}(q,t,E_{rec})|_{e^-} - \mathrm{Loss}(q,t,E_{rec})|_{\mu^-}
\end{equation}
Figure~\ref{fig:SKPIDScan} shows the distribution of the $e/\mu$ PID parameter variable for 12000 electron and muons, excluding those events missing a local minimum in the likelihood scan. While the full distribution extends well beyond the plotted range, we focus on the most interesting region, where the two populations cross over, to show the PID performance of our networks. The two particles types are well separated, with only a small fraction of events crossing the classification boundary at zero. A cluster of events around zero is due to events near the detector walls, which are more difficult to reconstruct as they tend to produce hits in a smaller number of PMTs. We note here that since this PID study is dnoe with all likelihood parameters except energy kept at their true values, the performance shown in this section should be taken as an indicative result. In a realistic event reconstruction setting, all the neural network input parameters will need to be estimated simultaneously by finding the global minimum in the negative log-likelihood surface.

\begin{figure}[h!]
\begin{center}
\includegraphics[width=\textwidth]{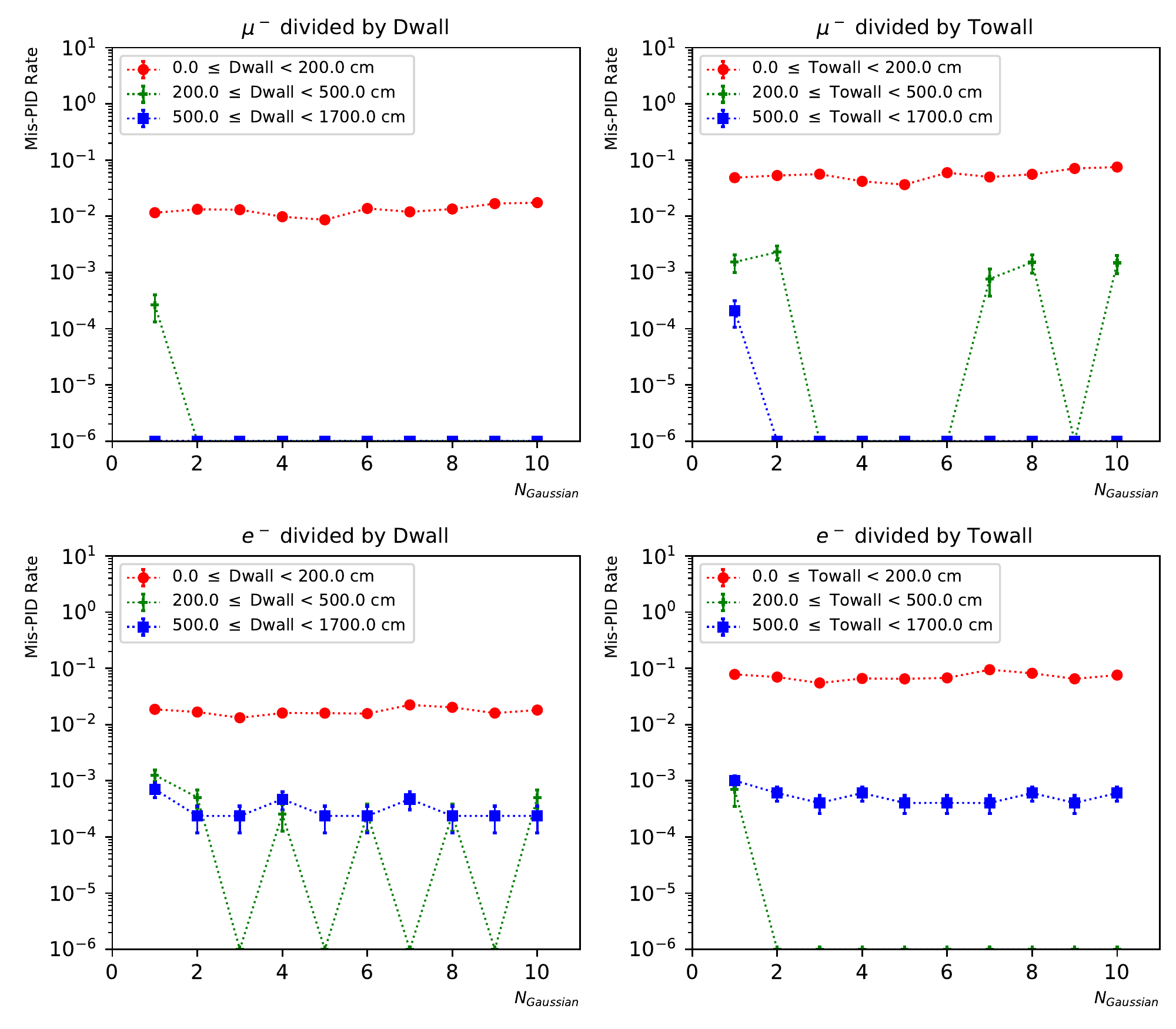}
\end{center}
\caption{Theparticle mis-identification rate of charge-only networks with 1$\sim$10 Gaussian from muon and electron events respectively. The three regions are 0$\sim$200, 200$\sim$500, and $\geq
$500 cm from the detector walls. The left column shows dependence of ``dwall" and the right ``towall". Statistical errors are also shown.}
\label{fig:multigaus_misid_vs_ngaus}
\end{figure}

\begin{figure}[h!]
\begin{center}
\includegraphics[width=\textwidth]{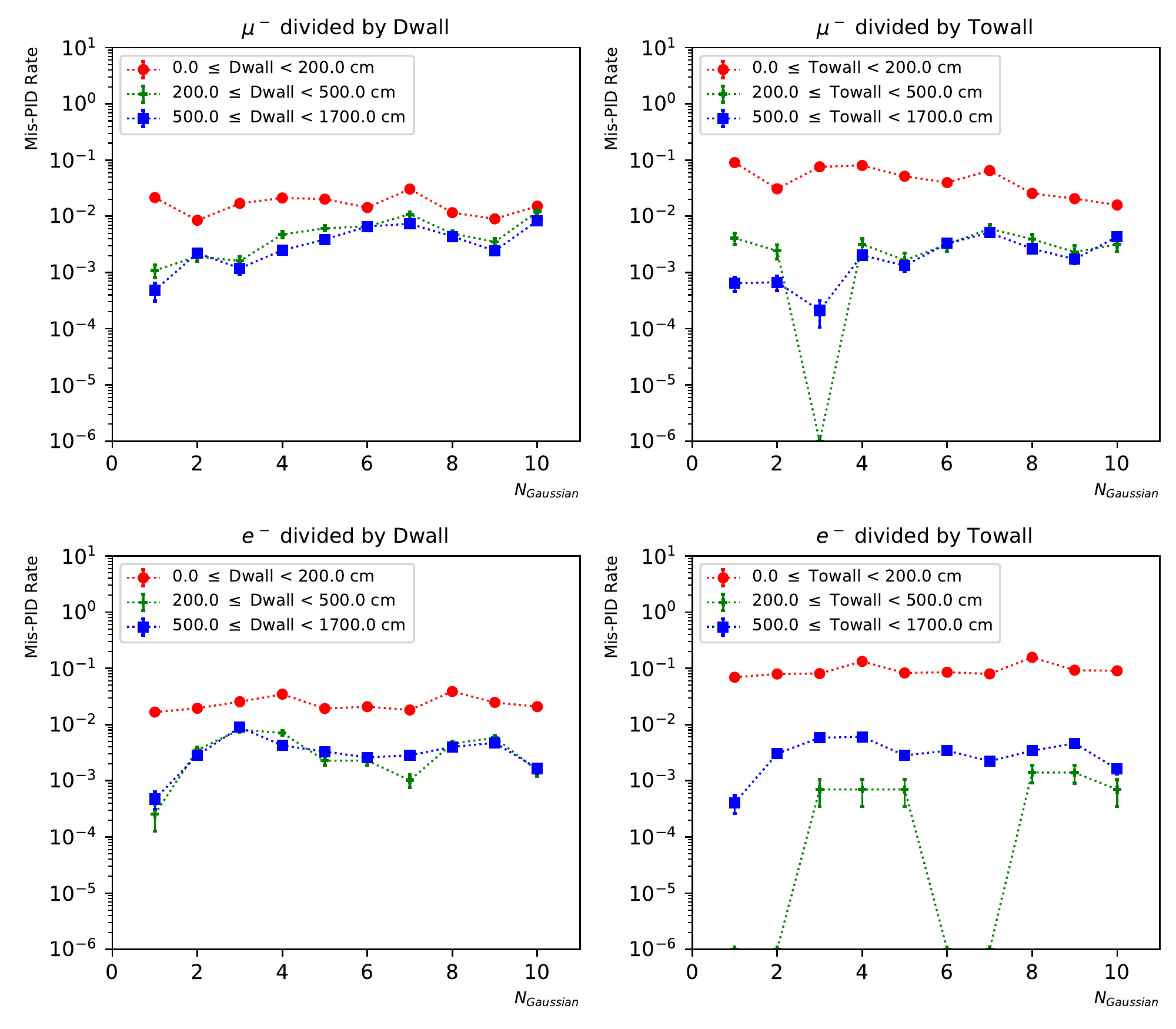}
\end{center}
\caption{Theparticle mis-identification rate of charge-time correlated networks with 1$\sim$10 Gaussian from muon and electron events respectively. The three regions are 0$\sim$200, 200$\sim$500, and $\geq
$500 cm from the detector walls. The left column shows dependence of ``dwall" and the right ``towall". Statistical errors are also shown.}
\label{fig:multigaus_qtcorr_misid_vs_ngaus}
\end{figure}

In Figure~\ref{fig:multigaus_misid_vs_ngaus} and \ref{fig:multigaus_qtcorr_misid_vs_ngaus} we present the particle mis-identification rate, defined as the wrong-sign $e/\mu$ PID fraction of each true particle type, with the events broken down in different detector regions. The same conditions as in Figures~\ref{fig:multigaus_ereso_vs_ngaus} and \ref{fig:multigaus_qtcorr_ereso_vs_ngaus} are applied. Both the ``dwall" and ``towall" parameters have strong impact to PID accuracy. In the charge-only networks, events sufficiently far from the walls show a noticeable improvement of PID accuracy with more Gaussian components used in the PMT charge response charge response approximation. Overall for the events reconstructed by the charge-only networks, the particle mis-identification rate is $<$ 0.1\% in the region of $\mathrm{dwall} \geq 200$ cm or $\mathrm{towall} \geq 200$ cm. On the other hand, this does not hold true for the charge-time correlated networks. In the ``near-wall" region, the charge-time correlated networks have comparable PID performance with the charge-only ones, but the central events have higher particle mis-identification rate by up to one order of magnitude and overall larger fluctuation is observed. Given the higher number of free parameters in a charge-time correlated network, this may indicate the lack of loss convergence and the need of more training. 

\section{Discussion}
In this section we discuss the results described above, as well as observations made during the development of the neural networks presented in this work.

\subsection{Neural network training}
While developing the neural network we faced training instability, with the loss function occasionally becoming not-a-number, particularly for more complex functional forms with several components in the Gaussian mixture. We achieved stability by carefully parameterizing the loss functions, as described above, and by pre-scaling our inputs such that they do not significantly exceed unity. Other attempts to stabilize the network training, such as the introduction of batch normalization layers were not as successful.

Especially for simpler versions of the neural networks, we observed that after a rapid decrease of the loss at the start of the training, it would stabilize at an intermediate value for a few hundred iterations before decreasing again in a short step to the stable minimum. Inspection of the network output during this period of metastability revealed a clear difference in the generated images before and after the short step, with localized features appearing suddenly after the step, while before the step the network produces a relatively uniform output for the entire image. After the network learns to localize the events, the quality of the generated rings gradually improves with training. This behaviour of the loss during training is significantly washed out when using more complex loss functions with several components.

\subsection{Network architecture studies}
\label{sec:PossibleMods}
We tested several modifications to the neural network architecture shown in Figure~\ref{fig:architecture}, while assessing the performance of the alternative architectures mostly by inspecting the behaviour of the loss function during training.

We studied the effects on the network performance of changing the number of fully connected layers and the number of nodes in each layer. These studies were done in a version of the network that generated data only for the barrel section of the detector. We changed the number of nodes in each fully connected layer from three to 1000 and found that 30 nodes were sufficient, and adding more nodes did not improve the results.  We found that at least two fully connected layers (excluding the input layer) were necessary to generate events on the barrel.

\begin{figure}[h!]
    \centering
    \includegraphics[width=0.5\textwidth]{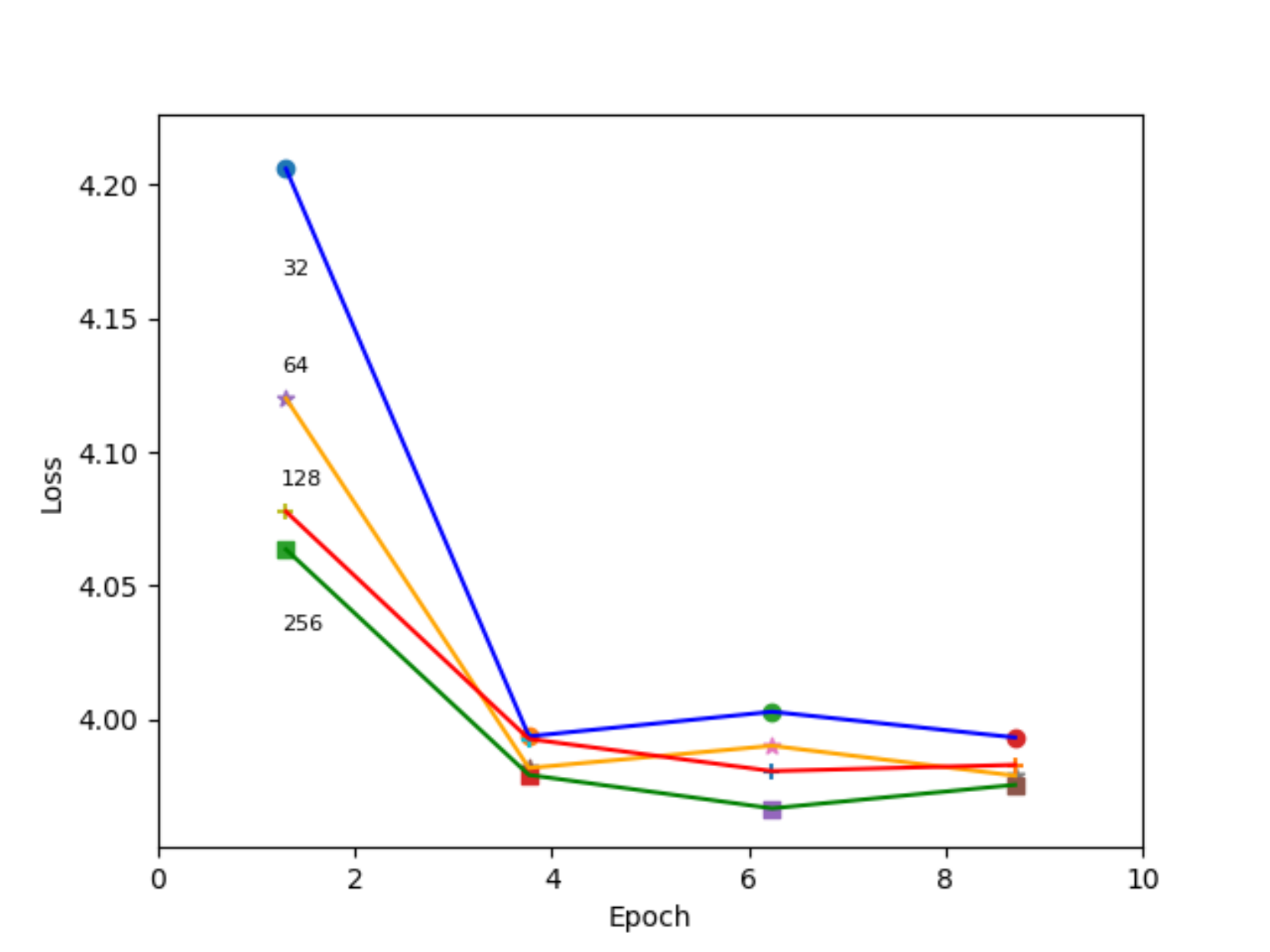}
    \caption{Loss as a function of training epoch for varying depths of the neural network's convolutional layers. The depth of the convolutional layers used for each training is indicated at the start of the respective line.}
    \label{fig:ChannelOpt}
\end{figure}

In addition, we studied the effect of the changing the depth of the network's convolutional layers using the single-Gaussian charge-only loss function. Depths of 32, 64, 128, and 256 were tested, where 64 is the default number. We found that increasing the depth  makes the network settle faster, while achieving a more or less similar loss value after training. An example of such a study is shown in Figure~\ref{fig:ChannelOpt}.

We also tested using three different activation functions throughout the network: \emph{ReLU}, \emph{tanh}, and \emph{LeakyReLU}. The \emph{LeakyReLU} activation function was examined with two different negative slopes, 0.1 and 0.5. The \emph{tanh} and \emph{LeakyReLU} performed equally well or worse compared to \emph{ReLU}.

\subsection{Network Reconstruction Performance}
\label{sec:nnperformance}
With the results shown in Section.\ref{sec:particlegun} and \ref{sec:likelihoodscan}, we notice that the multi-Gaussian PDF can improve the reconstruction of individual PMT responses, but has limited impact to the particle event reconstruction, possibly due to insufficient training. With the inclusion of PMT time responses, the charge-time correlated networks show larger uncertainty than the charge-only ones, especially for electrons. We have not determined what causes this and we will investigate this effect further.

\begin{figure}[h]
\begin{center}
\includegraphics[width=\textwidth]{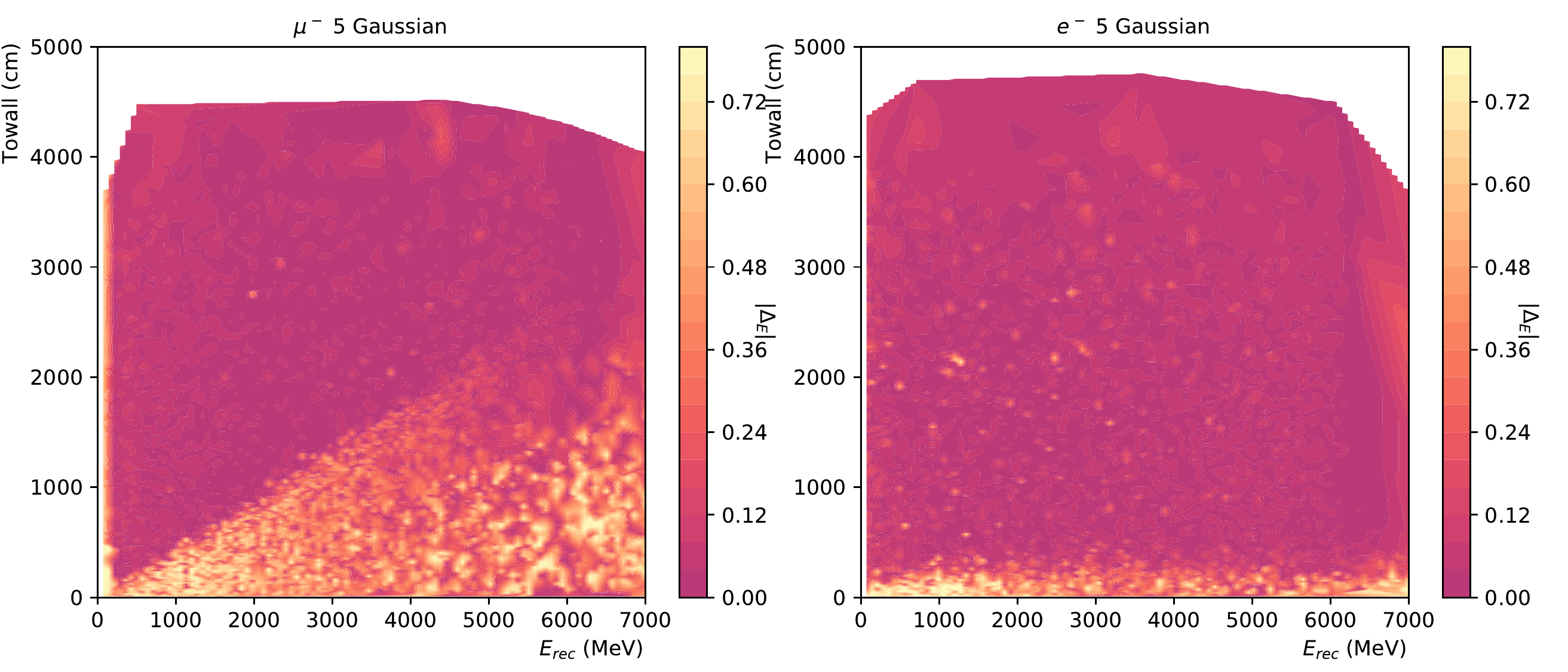}
\end{center}
\caption{The energy reconstruction performance of 12000 muon (left) and electron (right) events plotted against the $E_{rec}$ and towall for charge-only network with 5 Gaussian. The saturation level at 0.8 is from the energy range for likelihood scans.}
\label{fig:multigaus_mu_erec_vs_towall}
\end{figure}

Moreover, event reconstruction performance strongly depends on the event's location, as a near-wall one is less likely to deposit all of its energy in the detector, especially in the case of muons, and the Cherenkov ring is sampled by fewer PMTs for events with small ``towall". Figure~\ref{fig:multigaus_mu_erec_vs_towall} shows the distribution of $\Delta_{E}$ in the plane of $E_{rec}$ and ``towall" for muon and electron respectively. In the region where a particle's energy is fully captured by the detector, e.g. low $E_{rec}$ and high ``towall", the network shows excellent energy reconstruction accuracy. A linear relation is noticeable between muon's $E_{rec}$ and ``towall", which is consistent with the expectation: muon track lengths are approximately proportional to their energy and once a muon track escapes the detector it is no longer possible to accurately measure its energy. Water Cherenkov experiments typically include external detector elements to identify escaping events that cannot be reliably reconstructed. Electron showers, on the other hand, have a much more limited extent, and therefore their energy is accurately reconstructed even at high energies and relatively small values of ``towall". Another interesting observation in these results is the smooth appearance of well-reconstructed muon events at 100$\sim$200 MeV as shown in Figure~\ref{fig:multigaus_mu_erec_vs_towall}. These energies correspond roughly to the Cherenkov threshold and the smooth transition indicates the neural network is capable of identifying muon rings close to the threshold, albeit with a limited efficiency. Detecting particles close to their threshold is important particularly when considering events with multiple particle topologies, where very faint rings can coexist with much brighter rings, making their identification challenging.

\section{Conclusion}
We have developed a deep convolutional neural network that generates the likelihood function for the complete set of observables in a cylindrical water Cherenkov detector. We presented results using three likelihood function approximations using mixtures of Gaussian functions in one-dimension, taking into account only the hit charges, and in two-dimensions, taking into account both the hit charges and times. For the bivariate mixture model we explored two parameterizations of the Gaussian distribution, one which does not allow for correlations between charge and time in the individual components of the mixture, and one which includes this additional degree of freedom.

We have demonstrated that our neural networks are capable of accurately reproducing the features observed in the data and we found our results to be robust with respect to changes in the network architecture and loss function parameterizations, though longer training of our neural network will be necessary to confirm some of our conclusions.

We have studied the potential of these neural networks to be used in maximum likelihood reconstruction by scanning the loss as a function of the particle energy for a set of muon and electron events, and we demonstrated the loss is minimized close to the true particle energy. Taking the difference of between the minimum loss under the electron hypothesis and the minimum loss under the muon hypothesis to form a likelihood ratio test, the neural network as shown very promising particle identification performance.

The work presented here is an initial milestone in the path to achieve precise water Cherenkov event reconstruction using ML-based maximum likelihood estimation. Our future work will focus on the application of gradient descent algorithms to the neural network presented here, and further improving their performance based on event reconstruction metrics. We also plan to continue exploring alternative loss function parameterizations, such as mixtures of log-normal and other simple functions. We foresee the next major milestone in the project to be the extension of the neural network to multiple-particle topologies, which we expect to achieve by combining several single-particle likelihoods, possibly using ML techniques.

The approach presented here is not only promising in terms of potential improvements in reconstruction performance, but it is also easily applicable to any cylindrical water Cherenkov experiment, with extension to other geometries possible as long as they can be reasonably projected onto a set of two-dimensional images.


\section*{Acknowledgments}
We acknowledge the support of the Water Cherenkov Machine Learning (WatChMaL) group\footnote{https://www.watchmal.org/}, particularly in the early stages of the project. We would like to thank Stony Brook Research Computing and Cyberinfrastructure, and the Institute for Advanced Computational Science at Stony Brook University for access to the high-performance SeaWulf computing system, which was made possible by a \$1.4M National Science Foundation grant (\#1531492). This research was enabled in part by support provided by WestGrid (https://www.westgrid.ca/support/) and Compute Canada (www.computecanada.ca) with its high-performance heterogeneous cluster Cedar.

\bibliographystyle{unsrt}
\bibliography{arxiv_watercherenkov_generative_reco_bibliography}

\end{document}